# Competing Antiferromagnetic Phases in Multiferroic Wurtzite Transition-Metal Chalcogenides

Himanshu Mavani[†], Mohamed Elekhtiar[†], Kai Huang[†], Naafis Ahnaf Shahed, and Evgeny Y. Tsymbal*

*Department of Physics and Astronomy & Nebraska Center for Materials and Nanoscience, University of Nebraska, Lincoln, Nebraska 68588-0299, USA*

Antiferromagnetic (AFM) spintronics offers a pathway toward electrically controllable spin-based devices beyond ferromagnets. Here, we identify wurtzite Mn*X* (*X* = S, Se, Te) as a family of multiferroic materials hosting competing AFM phases, including altermagnetic, where nonrelativistic spin splitting can be controlled by ferroelectric polarization. Using density-functional theory and atomistic spin-model calculations, we show that all pristine Mn*X* compounds stabilize a stripe-type collinear AFM ground state, contrary to earlier predictions of an altermagnetic ground state, with the magnetic order governed by frustrated Heisenberg and biquadratic exchange interactions. We further demonstrate that Cr doping drives a transition to an A-type AFM phase that breaks Kramers spin degeneracy and realizes a g-wave altermagnetic state with large nonrelativistic spin splitting near the Fermi level. Importantly, this spin splitting can be deterministically reversed by polarization switching, enabling electric-field control of altermagnetic electronic structure without reorienting the Néel vector or relying on spin-orbit coupling. The close energetic proximity of the stripe AFM to a noncollinear all-in–all-out configuration indicates that wurtzite Mn*X* lies near a topological magnetic phase with finite scalar spin chirality, which may be stabilized by modest perturbations such as temperature, strain or chemical tuning. The distinct magnetic phases exhibit symmetry-selective linear and non-linear Hall responses, providing direct transport signatures of altermagnetism and polarization control. Together, these results establish doped wurtzite Mn*X* as a promising platform for altermagnet-ferroelectric multiferroics and electrically AFM spintronics.

*Keywords*: antiferromagnet, altermagnetic, ferroelectric, multiferroic

† These authors equally contributed to this work

* *E-mail*: tsymbal@unl.edu

## I. INTRODUCTION

Antiferromagnetic (AFM) spintronics represents a new frontier that could fundamentally reshape the landscape of spin-based technologies. Antiferromagnets offer distinct advantages over their ferromagnetic counterparts, including ultrafast spin dynamics operating in the terahertz regime, the absence of stray magnetic fields that allows for high-density device integration, and exceptional stability against magnetic perturbations [1,2]. These properties make AFM materials attractive for next-generation information storage and processing technologies, where speed, scalability, and robustness are critical.

To realize these advantages in practical devices, reliable detection and electrical readout of AFM order are essential. In conventional antiferromagnets, however, the absence of net magnetization renders such readout intrinsically challenging. This limitation can be overcome using altermagnets, a class of antiferromagnets characterized by alternating nonrelativistic spin splitting in their electronic band structures. These materials host spintronic phenomena such as the anomalous Hall effect, spin-polarized currents, and tunneling magnetoresistance, which were previously thought to be exclusive to ferromagnets [3-15].

A particularly intriguing class of AFM materials are multiferroics, which combine antiferromagnetism with spontaneous electric polarization, enabling direct coupling between electric and magnetic degrees of freedom [16,17]. In conventional collinear antiferromagnets, spin degeneracy is protected by the combined symmetry of space inversion and time reversal $PT$ or by a lattice translation combined with spin reversal $U\tau$. In non-centrosymmetric structures with coincident magnetic and crystallographic unit cells, both $PT$ and $U\tau$ symmetries are broken making such multiferroics favorable platform for realizing nonrelativistic spin-split antiferromagnets. However, if the magnetic ordering requires a supercell of the crystallographic unit cell, the $U\tau$ symmetry may remain preserved, thereby restoring spin degeneracy despite the absence of inversion symmetry [18]. Ferroelectric polarization provides an effective route to manipulate the sign of nonrelativistic spin splitting. In particular, studies have demonstrated that reversing electric polarization can switch the direction of spin splitting in collinear spin-split antiferromagnets, enabling electric-field control of spin-dependent electronic structures [19-23]. This establishes multiferroic antiferromagnets as a promising platform for electrically controllable spintronic functionalities without requiring manipulation of the Néel vector [24].

Hexagonal wurtzite-structured compounds (space group: $P6_3mc$) are compelling candidates for polar multiferroic antiferromagnets. Even in the absence of magnetism, these materials have recently attracted renewed interest owing to their large spontaneous polarization and the emerging possibility of electric-field-induced switching. Although conventional wurtzites are intrinsically polar, their polarization is generally non-switchable because of large reversal barriers. Nevertheless, the robust polarization along the hexagonal $c$ axis, often exceeding tens of $\mu C/cm^2$, persists to high temperatures and is accompanied by low dielectric losses, making these materials attractive for power and high-frequency electronics. A major breakthrough was the experimental demonstration of switchable ferroelectricity in wurtzite-structured nitrides, especially Sc- and Y-doped AlN [25-31], which established that polarization reversal can be realized within the wurtzite framework. Incorporating magnetic order into such polar wurtzite systems therefore offers a promising route to electrically tunable AFM spin splitting, opening new opportunities for spintronic functionality.

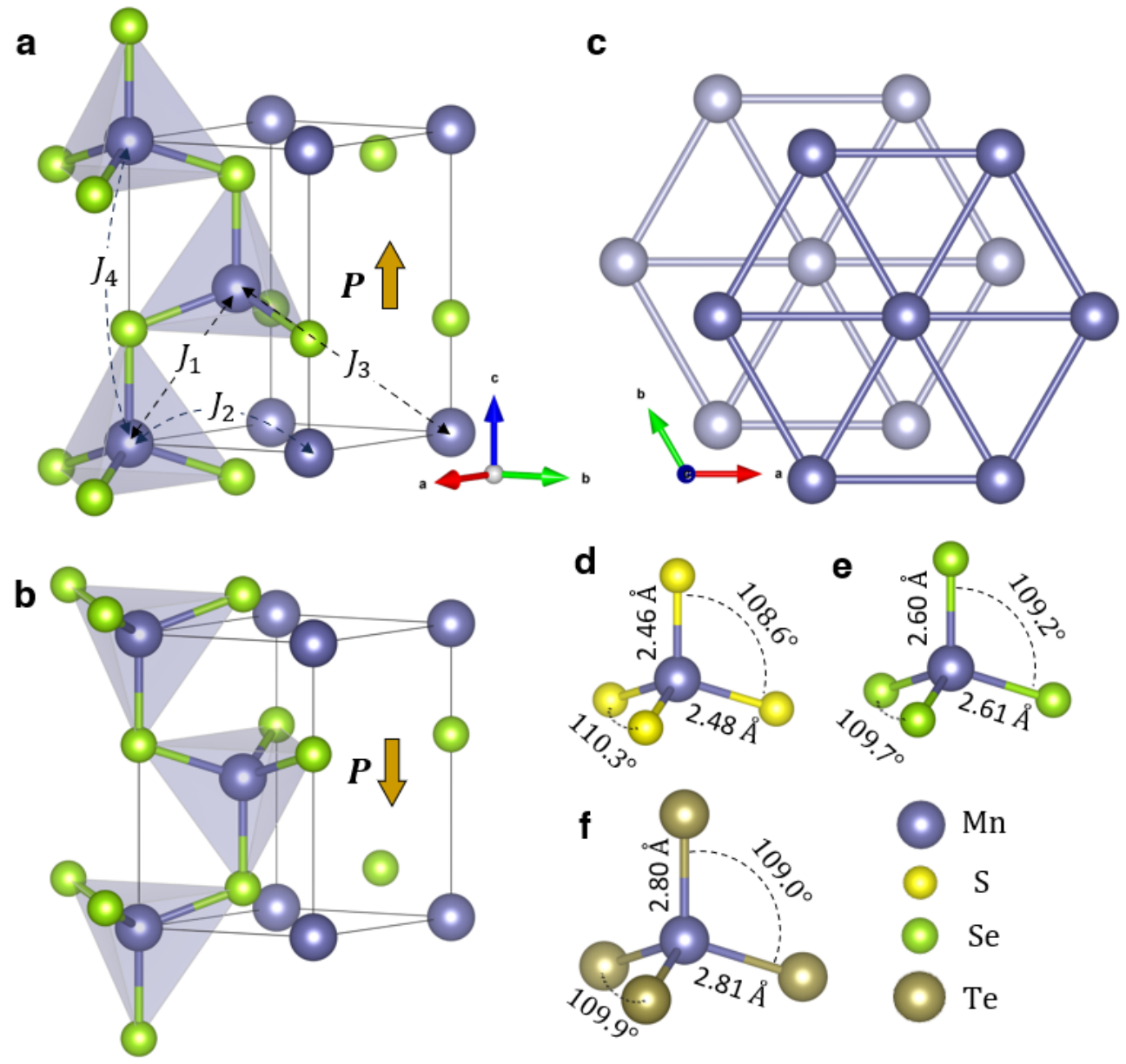

**FIG. 1**: Crystal structures of the polarization-up (a) and polarization-down (b) states in wurtzite Mn$X$ compounds. Dashed lines in (a) indicate first ($J_1$), second ($J_2$), third ($J_3$) and fourth ($J_4$) nearest-neighbor exchange interactions. (c) Top view of the Mn magnetic sublattice. (d–f) Tetrahedral geometry, bond angles, and bond lengths for the polarization-up state in MnS, MnSe, and MnTe, respectively.

Despite this potential, only a very limited number of wurtzite compounds are currently known to exhibit both ferroelectricity and magnetism [32-35]. Recent experimental advances have nevertheless demonstrated the realization of magnetic wurtzite phases, including the successful growth of wurtzite MnSe thin films by molecular beam epitaxy on GaAs(111) substrates with As-terminated surface using a CdSe/(Cd,Mg)Se buffer layer [36]. In addition, nanoscale wurtzite MnO has been realized through the thermal decomposition of $Mn(acac)_2$ on a carbon template [37]. Complementing these experimental developments, first-principles calculations predicted that wurtzite MnO and MnSe host an altermagnetic ground state [36,38,39].

Motivated by these findings, in this work, we perform a comprehensive theoretical investigation of the wurtzite family of multiferroic compounds. We focus on wurtzite materials that host competing collinear and noncollinear AFM phases, including the possibility of an altermagnetic state. For the latter, we demonstrate that the nonrelativistic spin splitting can be reversed by switching the electric polarization. Using first-principles density functional theory (DFT) in combination with atomistic spin models, we investigate the magnetic and ferroelectric properties of Mn$X$ ($X$ = S, Se, Te) single crystals, which have recently been identified as compounds that can be stabilized in the hexagonal wurtzite phase.

Contrary to the earlier predictions, our calculations show that all wurtzite Mn$X$ compounds stabilize in a stripe-type spin-degenerate AFM ground state. This conclusion is supported by a detailed analysis of the first four nearest-neighbor Heisenberg exchange interactions together with the nearest-neighbor biquadratic exchange and is further confirmed by atomistic spin-model simulations. Importantly, we find that the magnetic ground state in wurtzite Mn$X$ can be tuned by chemical substitution. In particular, Cr doping drives a transition from the stripe-type AFM order to an A-type AFM configuration, which breaks Kramers spin degeneracy and gives rise to an altermagnetic phase. Finally, we investigate the ferroelectric properties of the wurtzite Mn$X$ family. Using the Berry-phase approach, we predict a sizable spontaneous polarization ranging from 43 to 55 $\mu C/cm^2$ along the [0001] direction of the hexagonal lattice. The estimated coercive field of 1.8–2.1 $\mu C/cm^2$ falls within the range experimentally achieved in recently realized wurtzite ferroelectrics, indicating that polarization switching in wurtzite Mn$X$ should be experimentally accessible.

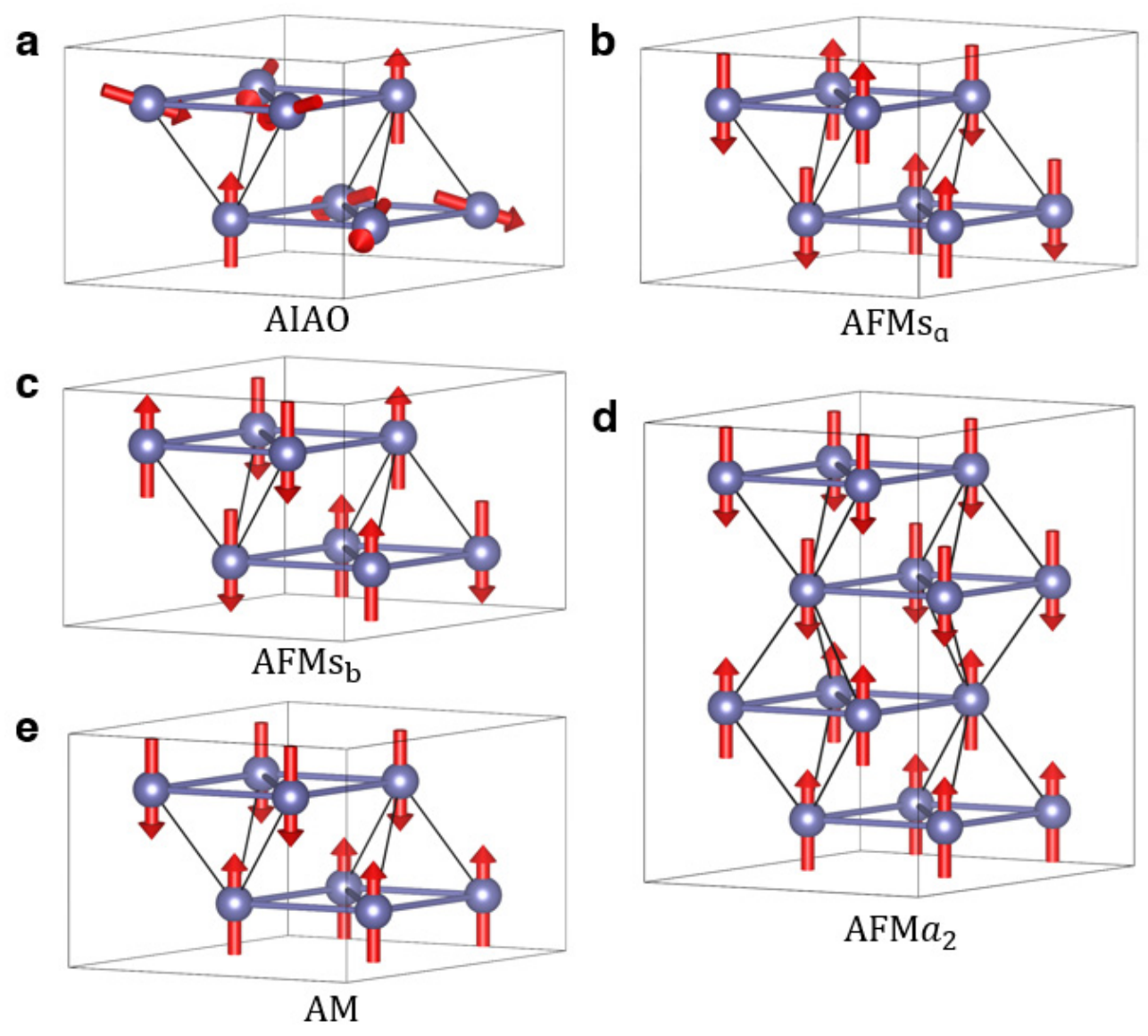

**FIG. 2**: Magnetic configurations of wurtzite Mn$X$: (a) non-collinear all-in-all-out (AIAO) phase; (b) stripe antiferromagnetic (AFM) phase $AFMs_a$; (c) stripe AFM phase $AFMs_b$; (d) block-type AFM phase $AFMa_2$; and (e) altermagnetic (AM) phase.

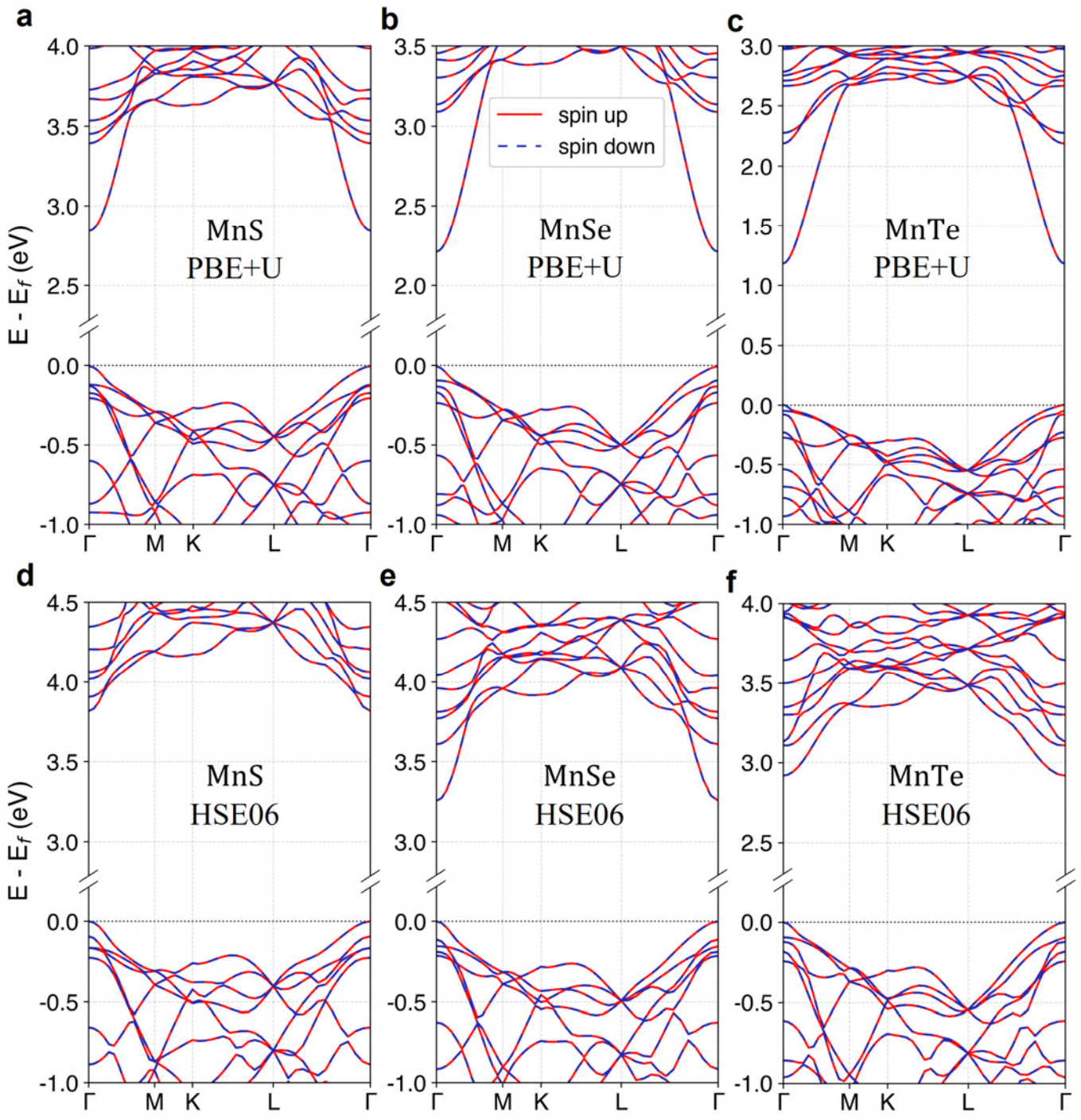

**FIG. 3**: Electronic band structures of AFMs$_\alpha$ phase in wurtzite Mn$X$ calculated using (a-c) PBE+U ($U$ = 4 eV) and (d-f) HSE06 methods. Spin-up (down) bands are indicated by solid red (dashed blue) curves.

## II. METHODS

First-principles DFT calculations were performed using both the Vienna *ab initio* Simulation Package (VASP) [40] and Quantum ESPRESSO (QE) [41]. Ion-electron interactions were treated using the projector augmented wave (PAW) in VASP [42,43] and optimized norm-conserving Vanderbilt pseudopotential in QE [44]. The Perdew-Burke-Ernzerhof (PBE) functional was used to treat the exchange-correlation energy [45]. A plane-wave cutoff energy of 500 eV and 1143 eV were set in VASP and QE, respectively. The magnetic ground state and doping effects were computed in QE with Hubbard $U$ values of 3 eV and 4 eV applied to the Mn $d$-orbitals [46]. To validate these results, the magnetic configurations were also studied using hybrid-functional HSE06 calculation in VASP [47,48]. For DFT+U calculations in QE, a $16 \times 16 \times 10$ k-point mesh was applied, whereas for HSE06 calculations in VASP utilized with $6 \times 6 \times 4$ mesh, and suitably reduced when supercells were considered. Atomic positions were relaxed until the force on each atom is less than $10^{-4}$ eV/Å, and self-consistency of electronic structure calculations is achieved with a tolerance of $10^{-8}$ eV. The resulting lattice parameters and band gaps for different wurtzite Mn$X$ compounds are listed in Table A1. The magnetic anisotropy energy $E_{MAE}$ was calculated using fully relativistic norm-conserving pseudopotentials in QE using a $24 \times 24 \times 15$ k-point mesh, which ensured convergence of $E_{MAE}$ to a few μeV. Cr doping was treated within the virtual crystal approximation (VCA) [49] by replacing the Mn sublattice with an effective virtual atom whose nuclear charge and electron count are linearly interpolated between Mn and Cr according to the doping level. Atomistic spin-model simulations for the magnetic ground of the Mn$X$ were performed in the Vampire code [50].

Ferroelectric properties were analyzed in VASP using PBE+U with $U$ = 4 eV on $8 \times 8 \times 5$ $k$-point grid. To simulate ferroelectric switching, the nudged elastic band method (NEB) was employed to identify a minimum energy path between the initial and final polarization states [51-53]. A climbing-image NEB scheme was used to calculate the energy profile for the domain-wall switching mechanism. In contrast, a fixed (non-variable) cell NEB approach was adopted to model uniform polarization reversal, thereby avoiding unphysical distortions of the bulk crystal lattice during coherent switching. A total of 15 intermediate states were calculated in each NEB calculation to provide the transition path. The convergence criterion for ionic relaxation was set to 0.001 eV/Å for all NEB calculations.

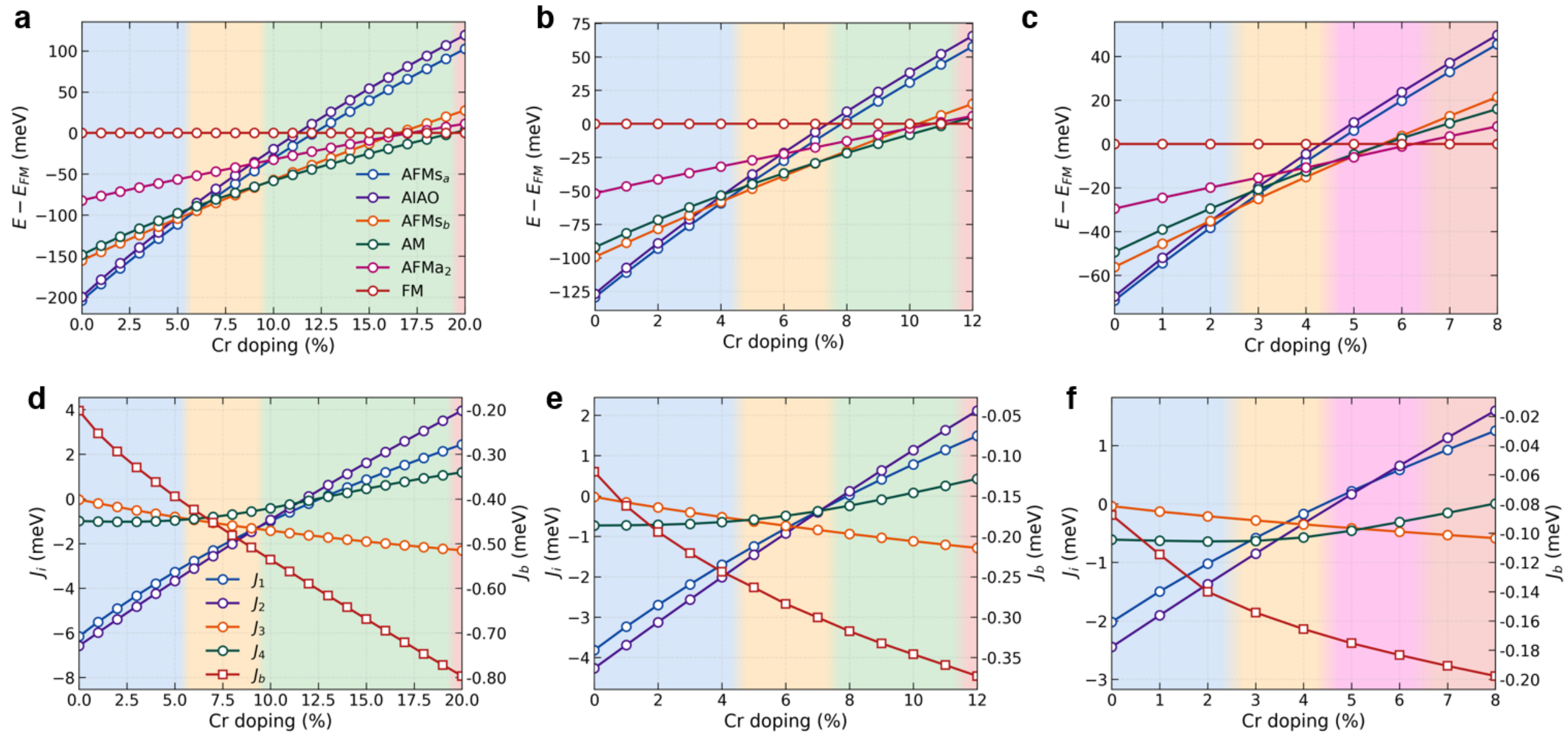

**FIG. 4**: (a-c) Energies of the magnetic configurations per non-magnetic unit cell, relative to the FM state, as a function of Cr doping in MnS (a), MnSe (b), and MnTe (c), computed using PBE+U ($U$ = 3 eV). (d-f) Corresponding exchange interactions and biquadratic exchange parameters as functions of Cr doping for MnS (d), MnSe (e), and MnTe (f). The blue, orange, green, pink and red shaded regions indicate doping intervals where AFMs$_\alpha$, AFMs$_b$, AM, AFMα$_2$ and FM are the lowest-energy configurations, respectively.

Several computational tools, including Bilbao Crystallographic Server [ 54 ], FindSpinGroup [ 55 ], and Pymatgen [ 56 ] were used for crystallographic analysis and electronic structure visualization.

## III. RESULTS

Wurtzite crystals belong to the non-centrosymmetric hexagonal space group $P6_3mc$. In Mn$X$ ($X$ = S, Se, Te) wurtzites, each Mn atom is tetrahedrally coordinated by $X$ atoms, forming polar Mn$X_4$ units that can orient either upward or downward along the $c$-axis. These two orientations correspond to polarization-up and polarization-down configurations (Fig. 1(a, b)).

### A. Magnetic properties

The magnetic sublattice of wurtzite Mn$X$ consists of Mn atoms arranged in AB-stacked two-dimensional (2D) triangular layers (Fig. 1(c)). To determine the magnetic ground state, we examine five AFM configurations and ferromagnetic (FM) configuration. The AFM cases include a noncollinear all-in–all-out (AIAO) configuration, two stripe-type AFM configurations (denoted as $\mathrm{AFMs_a}$ and $\mathrm{AFMs_b}$), block-type antiferromagnet ($\mathrm{AFM\alpha_2}$) and an altermagnetic (AM) configuration, as illustrated in Fig. 2.

The noncollinear AIAO magnetic phase is characterized by the arrangement of four Mn spins at the corners of each tetrahedron alternately pointing toward and away from the tetrahedral center. Consequently, the total spin moment of each Mn tetrahedron cancels out, yielding a net zero spin configuration. Notably, this AIAO phase possesses a finite spin chirality originating from its real-space spin arrangement, leading to the emergence of a topological Hall effect. This magnetic phase and the associated topological Hall effect have been experimentally observed in other material classes, such as $CoNb_3S_6$ and $CoTa_3S_6$, which feature similar magnetic lattices of Co atoms arranged in AB-stacked triangular layers [57].

The collinear stripe AFM states, $\mathrm{AFMs_a}$ and $\mathrm{AFMs_b}$, are composed of FM chains that alternate spin orientations between neighboring chains within one triangular layer. In the $\mathrm{AFMs_a}$ configuration, each Mn tetrahedron hosts two spin-up and two spin-down, ensuring net zero moment in each tetrahedron, similar to the AIAO phase. In contrast, $\mathrm{AFMs_b}$ breaks this balance, containing three spin pointing up and one pointing down or vice versa. Both stripe-type AFM phases preserve the combined $U\tau$ symmetry and therefore exhibit spin-degenerate band structures.

The fourth phase is a block-AFM configuration, $\mathrm{AFM\alpha_2}$, in which two consecutive layers are FM-aligned followed by two adjacent FM-ordered layers with opposite spin orientation. This phase also preserves the $U\tau$ symmetry. The fifth magnetic phase is a layered AFM configuration, commonly referred to as A-type antiferromagnet, where individual layers are FM-ordered but coupled antiferromagnetically along the stacking direction. In the wurtzite structure, this phase breaks $PT$ and $U\tau$ symmetries, leading to nonrelativistic spin-split electronic bands. This state realizes the AM phase, as discussed in detail in Sec. III(B).

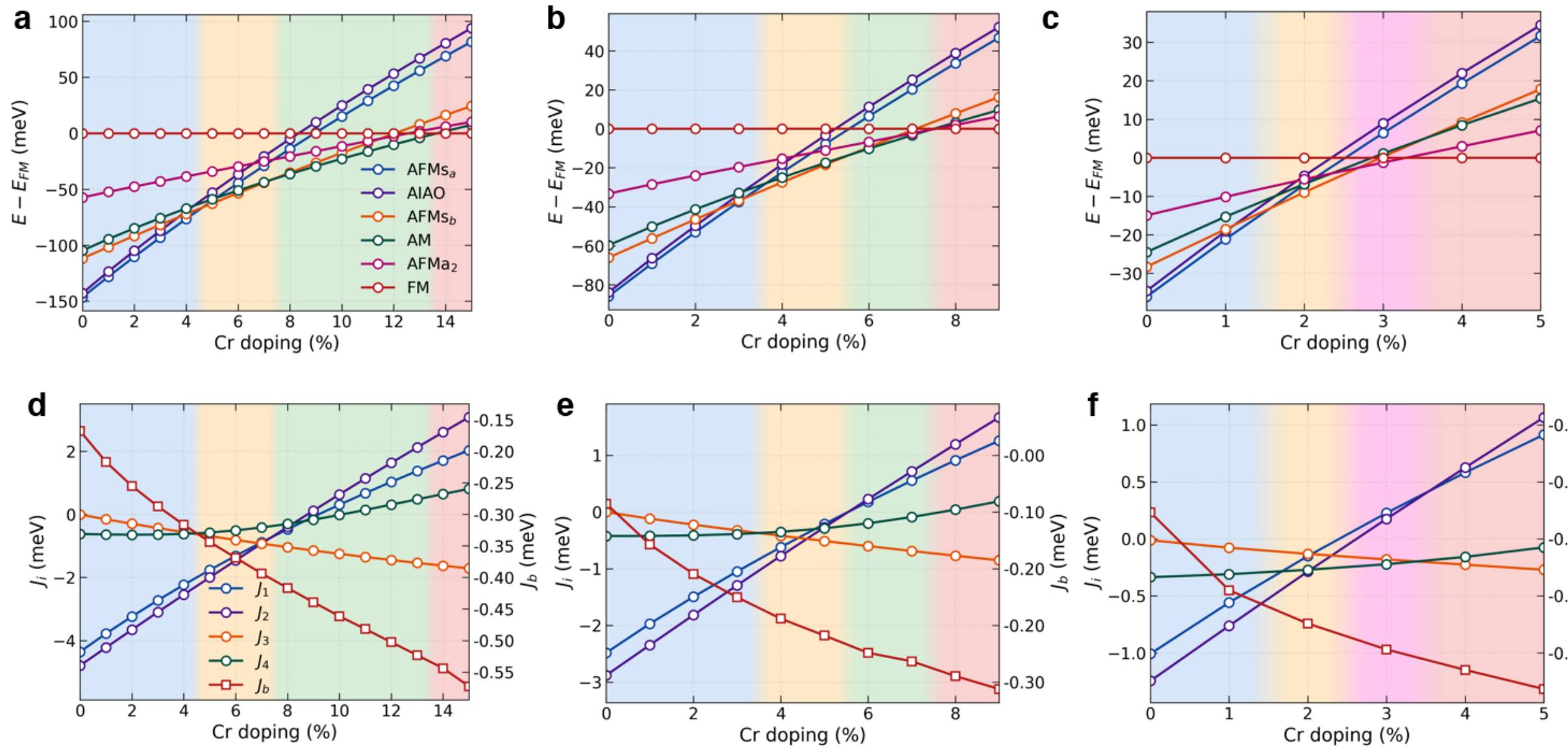

**FIG. 5**: (a-c) Energies of the magnetic configurations per non-magnetic unit cell, relative to the FM state, as a function of Cr doping in MnS (a), MnSe (b), and MnTe (c), computed using PBE+U ($U$ = 4 eV). (d-f) Corresponding exchange interactions and biquadratic exchange parameters as functions of Cr doping for MnS (d), MnSe (e), and MnTe (f). The blue, orange, green, pink and red shaded regions indicate doping intervals where $\mathrm{AFMs_a}$, $\mathrm{AFMs_b}$, AM, $\mathrm{AFM\alpha_2}$ and FM are the lowest-energy configurations, respectively.

The total energies of all magnetic configurations are evaluated using three computational schemes: PBE+U ($U$ = 3 eV), PBE+U ($U$ = 4 eV), and the hybrid functional HSE06. Across all methods, the relative energy sequence remains consistent for all three Mn$X$ ($X$ = S, Se, Te) wurtzite compounds, following the order

$$\mathrm{AFMs}_{\alpha} < \mathrm{AIAO} < \mathrm{AFMs}_{\mathrm{b}} < \mathrm{AM} < \mathrm{AFM}\alpha_2 < \mathrm{FM}.$$

Relative energies without spin-orbit coupling (SOC) are summarized in Table B1. The band structures corresponding to the lowest energy state, $\mathrm{AFMs}_{\alpha}$ phase, are presented in Fig. 3.

To understand the magnetic interactions, we employ a classical spin model incorporating Heisenberg exchange interactions up to the fourth nearest neighbors (NNs). Within this model, both the $\mathrm{AFMs}_{\alpha}$ and AIAO configurations remain degenerate in energy, even when exchange interactions are extended to any higher-range neighbors. Therefore, higher-order exchange terms, such as the biquadratic interaction, must be included to capture the small energy differences between these magnetic states found in our DFT calculations.

The spin-model Hamiltonian is therefore given by

$$H = -\sum_{n=1}^{4} J_n \sum_{\langle i,j\rangle_n} \boldsymbol{S}_i \cdot \boldsymbol{S}_j + J_b \sum_{\langle i,j\rangle_{1,2}} (\boldsymbol{S}_i \cdot \boldsymbol{S}_j)^2 \,,$$

where $J_1, J_2, J_3$, and $J_4$ denote the first, second, third, and fourth NN exchange interactions, respectively, as illustrated Fig. 1(a). The notation $\langle i,j\rangle_n$ indicates an $n^{th}$ NN spin pair $ij$, with each pair accounted twice in the summation, all the exchange interactions assumed to be isotropic. Positive (negative) $J_{ij}$ corresponds to FM (AFM) coupling. Because the first and second NNs are separated by comparable distances and the biquadratic exchange $J_b$ is relatively weak, we approximate $J_b$ to be identical for both shells. A negative (positive) $J_b$ favors collinear (non-collinear) spin configurations. The energies per crystallographic unit cell of the five magnetic configurations considered in Fig. 2 can then be expressed as

$$E_{\mathrm{AIAO}} = E_0 + 4J_1 + 4J_2 - 12J_3 - 4J_4 + 8/3\,J_b \,,$$
$$E_{\mathrm{AFMs}_{\alpha}} = E_0 + 4J_1 + 4J_2 - 12J_3 - 4J_4 + 24J_b \,,$$
$$E_{\mathrm{AFMs}_{\mathrm{b}}} = E_0 - 4J_1 + 4J_2 + 12J_3 - 4J_4 + 24J_b \,,$$
$$E_{\mathrm{AFM}\alpha_2} = E_0 - 12J_2 + 4J_4 + 24J_b \,,$$
$$E_{\mathrm{AM}} = E_0 + 12J_1 - 12J_2 + 12J_3 - 4J_4 + 24J_b \,,$$
$$E_{\mathrm{FM}} = E_0 - 12J_1 - 12J_2 - 12J_3 - 4J_4 + 24J_b \,,$$

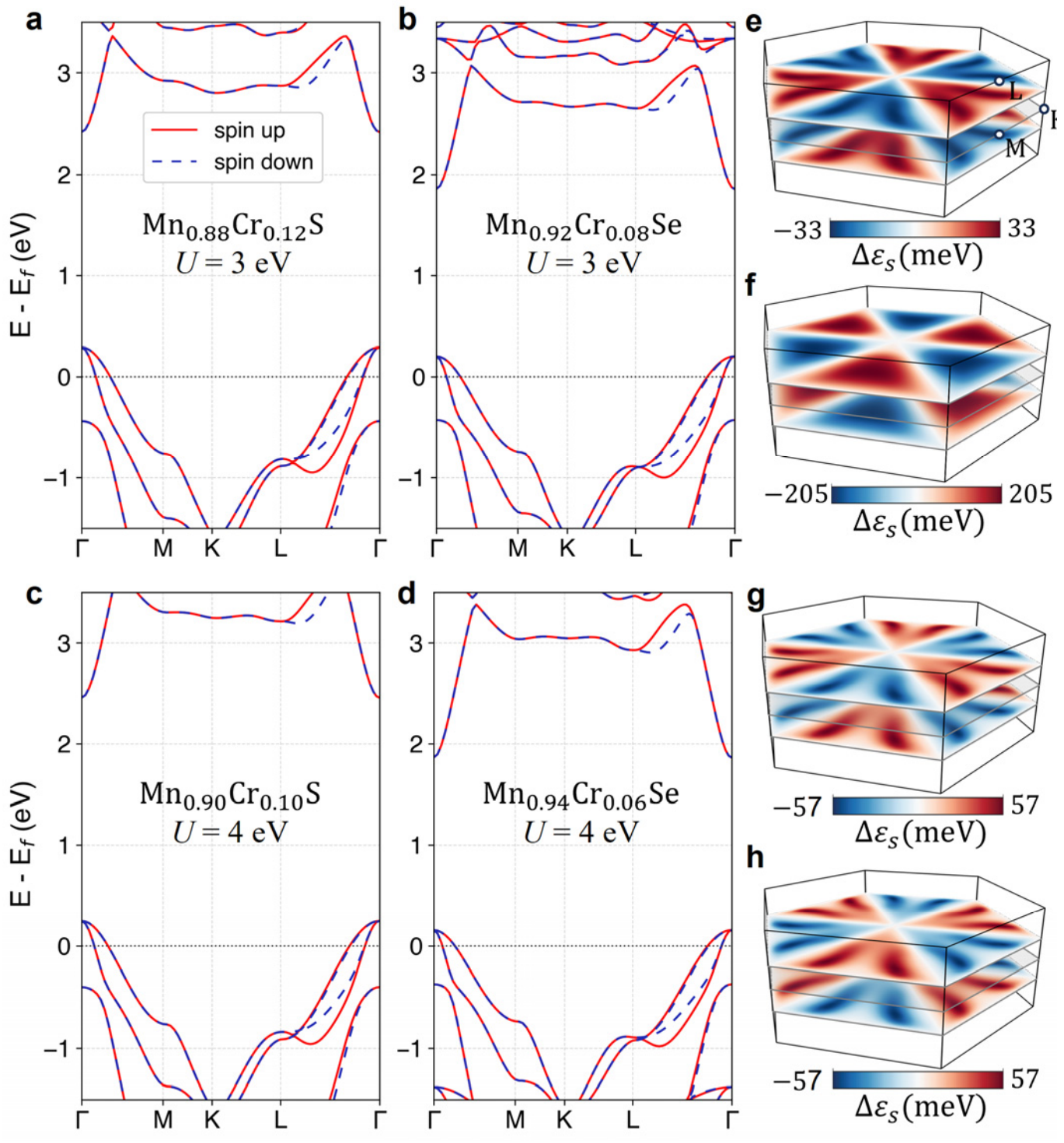

**FIG. 6**: (a-d) Electronic band structures of the altermagnetic phase in Cr doped MnS and MnSe wurtzite for $U$ = 3 eV and $U$ = 4 eV. Spin-up (down) bands are indicated by solid red (dashed blue) curves. (e, f) The non-relativistic spin-splitting energy, $\Delta\varepsilon_s(\boldsymbol{k}) = E_{\uparrow}(\boldsymbol{k}) - E_{\downarrow}(\boldsymbol{k})$, for the first (e) and second (f) bands crossing the Fermi level, evaluated in the $k_z = \pm 0.35\pi/c$ planes for $Mn_{0.88}Cr_{0.12}S$. (g, h) The spin-splitting energy of the first band crossing the Fermi level in the $k_z = \pm 0.35\pi/c$ planes for $Mn_{0.92}Cr_{0.88}Se$ in polarization-up (g) and polarization-down (h) states. Panels (e-h) are calculated for $U$ = 3 eV.

where $E_0$ is a non-magnetic contribution to the total energy and the spin vectors in the Hamiltonian are normalized to unity. The calculated exchange parameters are listed in Table B1. Both $J_1$ and $J_2$ are negative with comparable magnitudes, while $J_3$ and $J_4$ are small. The biquadratic exchange $J_b$ is weakly negative and therefore supports the $\mathrm{AFMs}_{\alpha}$ phase. Our atomistic simulations performed using these parameters conform $\mathrm{AFMs}_{\alpha}$ as the magnetic ground state.

Our results indicate that SOC plays only a minor role in determining the relative energetic stability of the magnetic configurations considered. As detailed in Appendix C, the magnetic anisotropy energy of wurtzite Mn$X$ is only a few tens of $\mu$eV (Table C1), which is well below the energy differences between the competing magnetic configurations listed in Table B1. As a result, the energetic hierarchy of the magnetic phases is governed primarily by exchange interactions. SOC mainly selects the preferred spin orientation, and for all compounds studied the $\mathrm{AFMs}_{\alpha}$ ground state favors an in-plane magnetic moment orientation, without altering the relative stability of the competing magnetic phases.

## B. Doping-induced magnetic phase evolution

Next, we explore the effect of doping on the AFM ground state. As shown in Table B1, the energies of the $AFMs_{\alpha}$ and AIAO phases nearly degenerate, and the energies of the $AFMs_{b}$ and AM phases are also comparable. This proximity suggests that carrier doping may provide an efficient route to manipulate the magnetic ground state.

Using VCA within the PBE+U framework, we explore hole doping by substituting Cr for Mn in the Mn$X$ wurtzite family. We find that, with increasing doping concentration, MnS and MnSe exhibit a systematic evolution of magnetic order, transitioning from $AFMs_{a}$ to $AFMs_{b}$, then to the AM phase, and ultimately to the FM state, as shown in Figs. 4 and 5. In MnTe, the overall trend persists, except that the AM phase is replaced by $AFMs_{\alpha}$. Increasing the Hubbard $U$ from 3 eV to 4 eV, reduces the stability windows of the magnetic phases, narrowing the range of doping concentrations over which each phase is stabilized.

The evolution of magnetic interactions is influenced by the chalcogen atom, with the overall exchange scale and phase stability window being largest in MnS and progressively reduced in MnSe and MnTe. As a result, MnS exhibits the widest stability range for distinct magnetic phases, whereas MnTe shows the narrowest. Upon increasing the Cr doping concentration, the exchange interactions display systematic trends that govern the evolution of the magnetic ground state. Specifically, both the NN in-plane and out-of-plane exchange, $J_1$ and $J_2$, decrease steadily in magnitude with the increasing Cr content, reflecting a progressive weakening in the dominant AFM exchange coupling. Near the transition between the $AFMs_{b}$ and AM phases, $J_1$ and $J_2$ change sign and become FM, with magnitude of $J_2$ remaining consistently larger than $J_1$. In contrast, the third NN exchange $J_3$ becomes increasingly negative (i.e., AFM) with doping and, within the AM regime, grows in magnitude to exceed $J_1$. This behavior identifies the second-NN out-of-plane coupling as the dominant AFM interaction, effectively enforcing AFM alignment between adjacent layers. When $J_1$ and $J_3$ become comparable in magnitude, the balance of interactions shifts, and the system ultimately stabilizes a FM ground state. The fourth-neighbor exchange $J_4$ also evolves with doping, changing sign from weakly AFM to weakly FM, although its magnitude remains small. Finally, the biquadratic exchange $J_b$ becomes increasingly negative with Cr substitution, consistently favoring collinear spin alignment across the entire doping range considered.

Focusing on the AM phase, we utilize the spin-space group formalism [6] to elucidate the symmetry origin of the nonrelativistic spin splitting that emerges in this configuration. A spin-space group symmetry operation is denoted as $[R_{\sigma}||R_i]$, where the operators to the left and right of the double bar act on the spin and real-space coordinates, respectively. The spin-space group of wurtzite Mn$X$ incorporates symmetries $[E||H] + [C_2||G-H]$ where $H$ is a subgroup of space group $G$ that includes $E$, $\pm C_{3z}$, $M_{[100]}$, $M_{[010]}$, and $M_{[110]}$ and $G-H$ includes $\pm\tau C_{6z}$, $\tau C_{2z}$, $\tau M_{[120]}$, $\tau M_{[1\bar{1}0]}$, and $\tau M_{[210]}$, with $\tau$ being half a unit cell translation along the $z$ direction. The symmetry operation $[C_2||C_{6z}\tau]$ transforms $\varepsilon(s, k_x, k_y, k_z)$ to $\varepsilon\left(-s, \frac{k_x}{2} - \frac{\sqrt{3}k_y}{2}, \frac{\sqrt{3}k_x}{2} + \frac{k_y}{2}, k_z\right)$. This operation, together with the spin-only symmetry $[\bar{C}_2||T]$ (where $\bar{C}_2$ is twofold spin rotation perpendicular to the collinear spin axis followed by spin inversion) of a collinear magnet, makes wurtzite Mn$X$ *g-wave* altermagnet.

These symmetry considerations are further substantiated by our DFT calculations for Cr-doped Mn$X$. Fig. 6(a-d) shows the electronic band structure of Cr doped MnS and MnSe calculated for two different values of the Hubbard $U$. It is evident that, independent of the doping level and the choice of $U$, the bands remain spin degenerate along the $\Gamma - \mathrm{M} - \mathrm{K} - \mathrm{L}$ path, whereas the spin degeneracy is lifted along the $\mathrm{L} - \Gamma$ path. This contrast arises from the spin-space symmetries of the AM phase. Along $\Gamma - \mathrm{M} - \mathrm{K} - \mathrm{L}$, the crystal momenta lie on symmetry lines that are invariant under a combined spin-space operation mapping $(s, \mathbf{k})$ onto $(-s, \mathbf{k})$ thereby enforcing spin degeneracy. In contrast, along $\mathrm{L} - \Gamma$ this protecting symmetry is absent, allowing a momentum-dependent nonrelativistic spin splitting.

In Figures 6(e-f), we plot the corresponding spin-splitting energy $\Delta\varepsilon_s(\boldsymbol{k}) = E_{\uparrow}(\boldsymbol{k}) - E_{\downarrow}(\boldsymbol{k})$ for two bands crossing the Fermi energy. The splitting alternates in sign across symmetry-related regions of the Brillouin zone and vanishes along symmetry-protected directions, forming a characteristic multi-lobed *g-wave* pattern [5]. For $Mn_{0.88}Cr_{0.12}S$, the spin-splitting magnitude reaches several tens of meV for the first band crossing, while it increases to about 200 meV for the second band crossing.

Importantly, nonrelativistic spin splitting in the AM phase of ferroelectric Mn$X$ can be fully reversed by switching the electric polarization. For a fixed Néel-vector orientation, the two opposite polarization states are related by the spin-space symmetry operation $[E||M_z]$, which maps $\varepsilon(s, k_x, k_y, k_z)$ onto $\varepsilon(s, k_x, k_y, -k_z)$. When combined with the g-wave AM symmetry, it enforces sign reversal of the spin splitting between the two polarization states, as illustrated in Figures 6(g-h) for $Mn_{0.88}Cr_{0.12}Se$. The electrically driven reversal of the **k**-dependent spin texture highlights the important functionality of ferroelectric altermagnets and opens a route toward nonvolatile, low-dissipation control of spin-polarized transport.

## C. Ferroelectric properties

Next, we explore the ferroelectric switching behavior of wurtzite Mn$X$ using the NEB method. As a reference case, we first consider a uniform polarization (UP) switching mechanism, in which the polarization of bulk Mn$X$ is assumed to reverse coherently from one polarization state to the opposite. Figures 7(a-c) show that the energy profile during switching exhibits two peaks separated by an energy valley. The calculated energy barriers are 256 meV/f.u. for MnS, 240 meV/f.u. for MnSe, and 247 meV/f.u. for MnTe. As illustrated in Figure 7(d) for MnSe, the switching proceeds via sequential displacements of the two Mn cations along the $+z$ direction within the unit cell, resulting in a reversal of the polarization from negative to positive. The intermediate energy minimum corresponds to the formation of a metastable quasi-2D Mn$X$ configuration [58], which gives rise to the characteristic double-barrier switching profile.

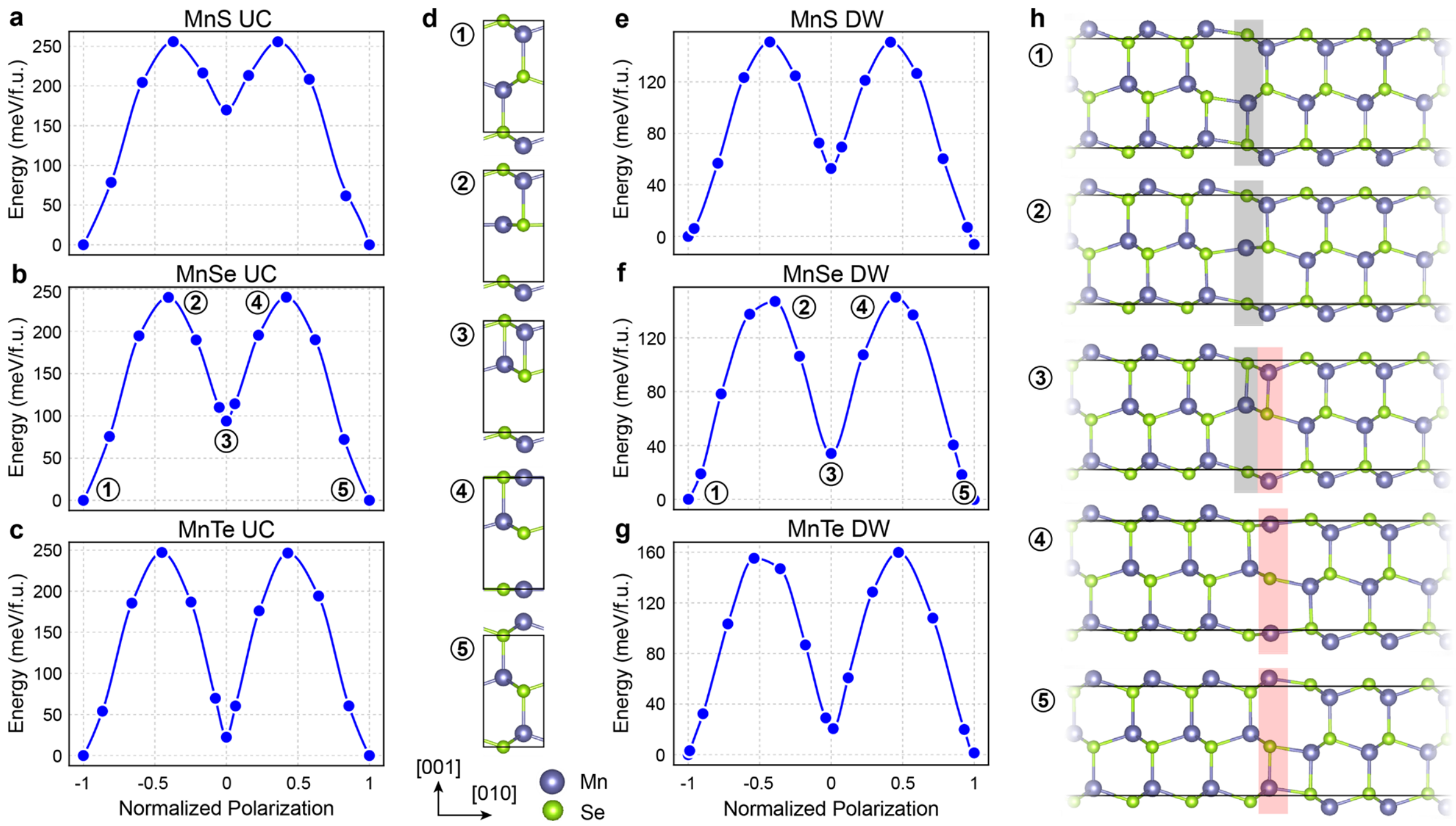

**FIG. 7**: Results of NEB calculations for polarization switching in ferroelectric Mn$X$ ($X$ = S, Se, Te). (a-c) Calculated energy profiles for uniform polarization (UP) switching in (a) MnS, (b) MnSe, and (c) MnTe. (d) Side-view of the selected intermediate states of MnSe along the path indicated in (b). (e-g) Calculated energy profile for domain-wall (DW) polarization switching in (e) MnS, (f) MnSe, and (g) MnTe. (h) The atomic structures of selected intermediate states of MnSe along the path indicated in (f), with the moving atoms highlighted by gray and light red backgrounds.

However, under practical conditions, UP switching is energetically unfavorable, as it requires the simultaneous reversal of polarization throughout the entire material. A more physically plausible switching mechanism involves the nucleation of reversed-polarization domains followed by their growth via domain-wall (DW) motion. To model the DW-mediated polarization switching, we construct a 1×8×1 supercell containing a DW configuration and perform NEB calculations. Figure 7(h) illustrates the DW motion for MnSe. Here, structure 1 represents a fully relaxed initial DW configuration separating two domains with polarization pointing up (left) and down (right). From structure 1 to structure 5, the DW translates rightward by one lattice constant, such that the final configuration is characterized by an expansion of polarization-up by one unit cell. The relaxed structures exhibit an atomically abrupt DW with the resulting configuration closely resembling the previously reported DW structure in wurtzite ZnO [59].

The atomic displacements associated with DW motion resemble those observed during UP switching. Initially, the cations and anions within the grey box in Figure 7(h) move he $+z$ and $-z$ directions, respectively, followed by analogous reflecting a sequential switching mechanism similar to that of the UP case. As a result, the energy profile exhibits two symmetric maxima separated by an intermediate energy minimum, resembling the UP-switching pathway. Notable, the energy barrier for MnSe DW motion is reduced to 150 meV/f.u., which can be attributed to the relaxation of surrounding atoms and the slight modulation of the unit-cell dimensions during switching. Similar DW-mediated switching is observed in MnS and MnTe, with calculated energy barriers of 151 meV/f.u. and 160 meV/f.u., respectively, as shown in Figures 7(e-g).

The coercive fields ($E_c$) associated with polarization switching are estimated from the steepest slopes of the energy-polarization curves shown in Figures 7(e-g), corresponding to the maximum value of $dE/dP$. Here, $E$ denotes the total energy and $P$ the polarization, which is approximated from atomic displacements weighted by the Born effective charges listed in Table D1. The estimated coercive field for MnSe is $E_c \approx 2.1$ MV/cm, while the corresponding values for MnS and MnTe are 1.8 MV/cm and 1.9 MV/cm, respectively. These values are comparable to the coercive fields reported for wurtzite ferroelectrics, such as AlN and Sc-doped AlN, where polarization switching typically occurs at electric fields of order $1-3$ MV/cm (e.g., [28]). This comparison indicates that polarization switching in wurtzite Mn*X* should be experimentally accessible.

## IV. DISCUSSION

Two-dimensional triangular magnetic lattices are well known to host a wide variety of frustrated magnetic states, including noncollinear antiferromagnetism and quantum spin-liquid phases, and have therefore been studied extensively in the literature [60]. In bulk materials, most prior work has focused on ABC-stacked triangular lattices, as this stacking sequence is realized in a large number of naturally occurring compounds. By contrast, AB-stacked triangular magnetic layers in bulk form remain relatively unexplored [ 61 ]. The present study of competing AFM configurations and predicted exchange interactions in wurtzite Mn*X* which features AB-stacked Mn triangular layers, therefore establishes a new materials platform for investigating frustrated magnetism in three dimensions.

Given the wide range of previous theoretical predictions for wurtzite Mn*X*, including FM, stripe-type AFM, noncollinear, and AM ground states [36,38,39,62-65], a comprehensive and systematic analysis is essential. Our results show that the collinear stripe AFM $AFMs_{\alpha}$ is the ground in pristine Mn*X* but lies energetically close to the noncollinear AIAO configuration. A similar energetic competition is observed in $CoTe_3S_6$, which also contains AB-stacked triangular layers of magnetic ions and stabilizes an AIAO phase at low temperature, while thermal effects drive a transition to a collinear state [66,67]. In wurtzite Mn*X*, this near-degeneracy suggests the opposite possibility: thermal tuning, strain, or magnetic-field cooling may stabilize the AIAO phase, opening a potential route to realizing real-space topological magnetism in a polar non-centrosymmetric host.

Our results predict that Cr doping provides an additional and experimentally viable control parameter. Beyond driving transitions between distinct AFM phases, Cr substitution shifts the Fermi level into the valence band, rendering the system conducting (semiconducting). The non-zero conductivity enables direct experimental identification of magnetic phase changes via electronic transport measurements, including the linear anomalous Hall effect (AHE) and symmetry-allowed higher-order nonlinear responses [68-71]. As detailed in Table E1, the distinct magnetic phases considered here support qualitatively different linear and second-order conductivity tensors associated with the Berry curvature dipole (BCD) and the quantum metric dipole (QMD), providing clear experimental fingerprints of both magnetic symmetry and polarization state. In particular, the linear AHE is symmetry-allowed in the AIAO phase and in the AM phase (the latter for $\hat{m} \parallel [1\bar{1}0]$), while it is strictly forbidden in the other AFM phases considered. Second-order nonlinear Hall responses associated with the QMD are also allowed in the AIAO and AM phases, whereas in the remaining AFM phases only BCD-driven nonlinear contributions are permitted.

Importantly, these linear and nonlinear transport responses exhibit distinct behavior under polarization reversal. As summarized in Table E2, depending on the magnetic moment orientation, specific components of the conductivity tensor change sign upon switching the ferroelectric polarization. This symmetry-controlled sign reversal provides an electrical means to detect the Néel vector orientation in these compounds, provided that ferroelectric polarization in conducting MnX remains switchable. In the semiconducting regime, nonlinear photocurrent generation can be employed to distinguish AM order from conventional AFM states. In particular, it has been proposed that the broken inversion symmetry in these compounds enables a second-order normal injection current that is directly correlated with the AM spin-splitting energy [72].

Taken together, these results establish doped wurtzite MnX as a promising materials platform for exploring frustrated magnetism, altermagnetism, and electrically tunable spin-dependent transport within a single material family.

## V. CONCLUSIONS AND OUTLOOK

In conclusion, we have presented a comprehensive theoretical investigation of competing AFM phases in multiferroic wurtzite Mn*X* (*X* = S, Se, Te), revealing a rich magnetic phase landscape arising from frustrated interactions on AB-stacked triangular Mn lattices. Contrary to earlier predictions, we demonstrate that the magnetic ground state of pristine wurtzite Mn*X* is a stripe-type collinear antiferromagnet that remains spin degenerate. This result is robust across multiple first-principles approaches and is quantitatively explained by a detailed analysis of Heisenberg and biquadratic exchange interactions, further validated by atomistic spin-model simulations.

Importantly, we identify chemical substitution as an effective and experimentally viable route to engineer alternative AFM phases, including altermagnetism, within this material family. Specifically, we predict that Cr doping induces a controllable transition from stripe-type AFM order to an A-type AFM configuration that breaks Kramers spin degeneracy and realizes a g-wave AM state. In this regime, we observe significant nonrelativistic spin splitting, reaching tens to hundreds of meV near the Fermi level, whose momentum dependence is dictated by spin-space symmetries. Crucially, we show that this spin splitting can be deterministically reversed by switching the ferroelectric polarization, enabling electric-field control of AM band structures without reorienting the Néel vector or relying on spin-orbit coupling. This mechanism represents a distinctive form of magnetoelectric coupling unique to ferroelectric altermagnets.

Beyond collinear antiferromagnetism, our results reveal a close energetic proximity of the stripe AFM ground state to the noncollinear AIAO configuration, indicating that that wurtzite MnX resides near a topological magnetic instability. As a result, modest perturbations, such as temperature, epitaxial strain, magnetic-field cooling, or chemical tuning, may be sufficient to stabilize the AIAO phase. Because AIAO order carries finite scalar spin chirality, its realization in wurtzite Mn*X* would enable real-space topological magnetism, potentially giving rise to anomalous and topological Hall responses even in the absence of net magnetization. The polar non-centrosymmetric crystal structure further distinguishes this scenario from previously studied AIAO systems, offering additional routes for electrical control.

Our symmetry analysis further establishes clear experimental signatures of the different magnetic phases through their linear and nonlinear transport responses. As detailed in Appendix E, only specific phases, namely the AIAO and selected AM configurations, support a linear anomalous Hall effect, while other AFM phases exhibit distinct second-order responses governed by Berry curvature and quantum metric dipoles. Importantly, subsets of these conductivity tensor components change sign upon polarization reversal, providing an electrical means to detect both magnetic symmetry and Néel-vector orientation, provided that ferroelectric switching remains operative in the conducting regime. This establishes transport measurements as a powerful probe of magnetic order and magnetoelectric functionality in doped wurtzite Mn*X*.

Our ferroelectric analysis further supports the viability of wurtzite Mn*X* as a multiferroic system. The predicted spontaneous polarizations (~40–60 μC/cm²), moderate coercive fields (~1–2 MV/cm), and DW-mediated switching pathways indicate that polarization reversal should be experimentally accessible, particularly in thin films or doped systems where domain-wall motion dominates the switching dynamics. The atomically abrupt domain walls and their close resemblance to those found in established wurtzite ferroelectrics such as AlN and ZnO suggest favorable kinetics and structural stability for device-relevant operation.

Looking forward, our results position Cr-doped wurtzite Mn*X* as a promising realization of altermagnet-ferroelectric multiferroics, in which magnetoelectric coupling operates through symmetry-controlled nonrelativistic spin splitting rather than conventional spin-orbit mechanisms. Experimentally, epitaxial growth of doped wurtzite Mn*X* thin films followed by transport measurements, such as linear and nonlinear Hall effects or tunneling magnetoresistance, could directly probe the predicted AM signatures and their electric-field reversibility. From a theoretical perspective, finite-temperature effects, epitaxial strain, and interface engineering merit further investigation, as they may stabilize additional competing magnetic phases or enhance magnetoelectric responses. More broadly, this work establishes a general design strategy for electrically controllable antiferromagnetic spintronics, unifying frustrated magnetism, altermagnetism, and ferroelectricity within a single, symmetry-driven materials framework.

## ACKNOWLEDGMENTS

The authors thank Stefan Blügel for useful discussions. This work was supported by the National Science Foundation (NSF) through the EPSCoR RII Track-1 Program under Award No. OIA-2044049 (first-principles modeling of novel antiferromagnets) and Division of Materials Research under Award No. DMR-2316665 (atomistic spin-dynamics simulations). Theoretical modeling of ferroelectric properties of wurtzite antiferromagnets was supported by the U.S. Department of Energy, Office of Science, Office of Basic Energy Sciences, through the DOE EPSCoR program under Award No. DE-SCSC0026103. Computations were performed at the University of Nebraska Holland Computing Center.

## DATA AVAILABILITY

The data that support the findings of this paper are not publicly available. The data are available from the authors upon reasonable request.

## APPENDIX A: LATTICE PARAMETERS AND BAND GAPS

**Table A1.** Calculated lattice parameters and band gaps of wurtzite Mn*X* structures.

| Compound | Method | $U$ (eV) | $a$ (Å) | $c$ (Å) | $E_g$ (eV) |
|---|---|---|---|---|---|
| MnS | PBE+U | 3 | 4.019 | 6.434 | 2.84 |
| | PBE+U | 4 | 4.035 | 6.463 | 2.85 |
| | HSE06 | – | 4.022 | 6.420 | 3.82 |
| MnSe | PBE+U | 3 | 4.203 | 6.813 | 2.20 |
| | PBE+U | 4 | 4.245 | 6.845 | 2.22 |
| | HSE06 | – | 4.193 | 6.797 | 3.26 |
| MnTe | PBE+U | 3 | 4.544 | 7.386 | 1.29 |
| | PBE+U | 4 | 4.587 | 7.427 | 1.18 |
| | HSE06 | – | 4.534 | 7.327 | 2.92 |

## APPENDIX B: MAGNETIC ENERGIES AND EXCHANGE INTERACTIONS

**Table B1.** Energies of the AFM configurations considered in Fig. 2 relative to the FM reference state, and the extracted exchange constants. Energies are expressed in meV per crystal unit cell, while exchange values are in meV. Spins are normalized to unity.

| Compound | Method | $U$ (eV) | Magnetic moment ($\mu_B$) | $E_{\mathrm{AFMs_a}}$ (meV /u.c.) | $E_{\mathrm{AIAO}}$ (meV /u.c.) | $E_{\mathrm{AFMs_b}}$ (meV /u.c.) | $E_{\mathrm{AFMa_2}}$ (meV /u.c.) | $E_{\mathrm{AM}}$ (meV /u.c.) | $J_1$ (meV) | $J_2$ (meV) | $J_3$ (meV) | $J_4$ (meV) | $J_b$ (meV) |
|---|---|---|---|---|---|---|---|---|---|---|---|---|---|
| MnS | PBE+U | 3 | 4.27 | -203.77 | -199.45 | -155.18 | -82.11 | -148.35 | -6.155 | -6.581 | -0.027 | -0.992 | -0.202 |
| | PBE+U | 4 | 4.33 | -146.39 | -142.81 | -111.71 | -57.36 | -104.74 | -4.357 | -4.792 | -0.007 | -0.624 | -0.168 |
| | HSE06 | – | 4.48 | -216.27 | -211.78 | -164.32 | -87.15 | -157.82 | -6.555 | -6.962 | -0.020 | -1.030 | -0.211 |
| MnSe | PBE+U | 3 | 4.39 | -129.32 | -126.77 | -99.30 | -51.89 | -92.11 | -3.817 | -4.266 | -0.021 | -0.730 | -0.120 |
| | PBE+U | 4 | 4.44 | -85.85 | -83.87 | -66.05 | -33.24 | -59.69 | -2.484 | -2.881 | -0.003 | -0.425 | -0.093 |
| | HSE06 | – | 4.48 | -141.19 | -138.26 | -106.94 | -59.50 | -105.48 | -4.367 | -4.458 | -0.028 | -0.845 | -0.137 |
| MnTe | PBE+U | 3 | 4.50 | -71.44 | -69.57 | -56.19 | -29.55 | -49.35 | -2.018 | -2.446 | -0.038 | -0.610 | -0.088 |
| | PBE+U | 4 | 4.56 | -36.08 | -34.57 | -28.31 | -14.94 | -24.50 | -1.008 | -1.247 | -0.012 | -0.337 | -0.071 |
| | HSE06 | – | 4.49 | -100.54 | -98.85 | -77.78 | -40.10 | -69.92 | -2.896 | -3.387 | -0.017 | -0.643 | -0.079 |

## APPENDIX C: MAGNETIC ANISOTROPY

The magnetic anisotropy energy (MAE), defined as $E_{MAE} = E_{\boldsymbol{m}||\boldsymbol{a}} - E_{\boldsymbol{m}||\boldsymbol{c}}$, is calculated for the wurtzite MnX compounds following the procedure described in Sec. II. We find that all compounds exhibit an in-plane easy axis, with negative MAE values on the order of a few μeV/u.c., as summarized in Table C1. In Figure C1, we plot the total energy $E$ of MnSe as the magnetic moments are rotated by an angle $\theta$ away from the $c$ axis, referenced to the energy $E_c$ for moments aligned along the $c$ axis. The angular dependence of energy is well described by a uniaxial form, $E - E_c = E_{MAE} \sin^2\theta$ . Similar behavior is obtained for MnS and MnTe.

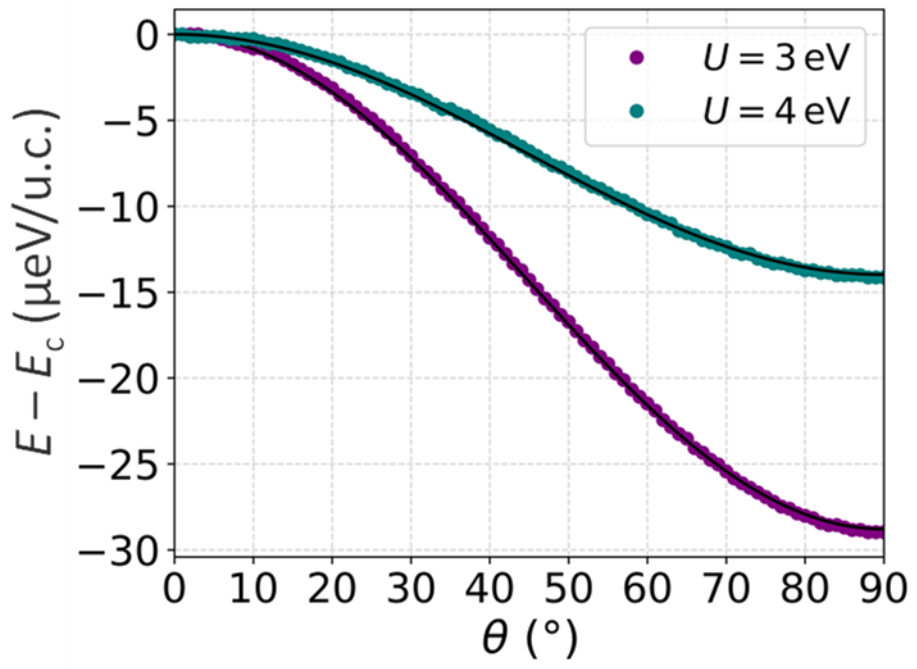

**FIG. C1**: Total energy of wurtzite MnSe as a function of the angle $\theta$ between the magnetic moments and the $c$ axis, referenced to the energy $E_c$ of moments aligned along the $c$ axis. The solid line represents a fit to the uniaxial anisotropy form $E - E_c = E_{MAE}\sin^2\theta$.

**Table C1.** The magnetic anisotropy energy $E_{MAE} = E_{\boldsymbol{m}||\boldsymbol{a}} - E_{\boldsymbol{m}||\boldsymbol{c}}$ in units of μeV/u.c. for the Hubbard $U$ = 3 eV and $U$ = 4 eV.

| Compound | $U$ = 3 eV | $U$ = 4 eV |
|---|---|---|
| MnS | -37.8 | -24.1 |
| MnSe | -28.8 | -14.0 |
| MnTe | -85.3 | -31.0 |

## APPENDIX D: POLARIZATION AND NEB CALCULATIONS

The polarizations of wurtzite structures were calculated as shown in Figure D1. During the ferroelectric polarization reversal from the $P_{up}$ to $P_{down}$ state, discontinuities arising from the multivalued nature of polarization were removed the standard Berry-phase branch-choice procedure [ 73 ]. The resulting polarization values are summarized in Table D1. The systematic increase in polarization from MnTe to MnS can be attributed to the increasing electronegativity of the lighter chalcogen elements, which enhances the ionic character of the bonding. As a result, cation displacements produce a larger change in polarization, despite the concomitant decrease in unit-cell volume for lighter chalcogens.

To validate our methodology, we calculated the spontaneous polarization of AlN to be 133.7 μC/cm$^2$, in good agreement with previous first-principles results [76]. The polarization of ZnO is obtained as 81.2 μC/cm$^2$, which is close to the reported value of approximately 100 μC/cm$^2$ [74].

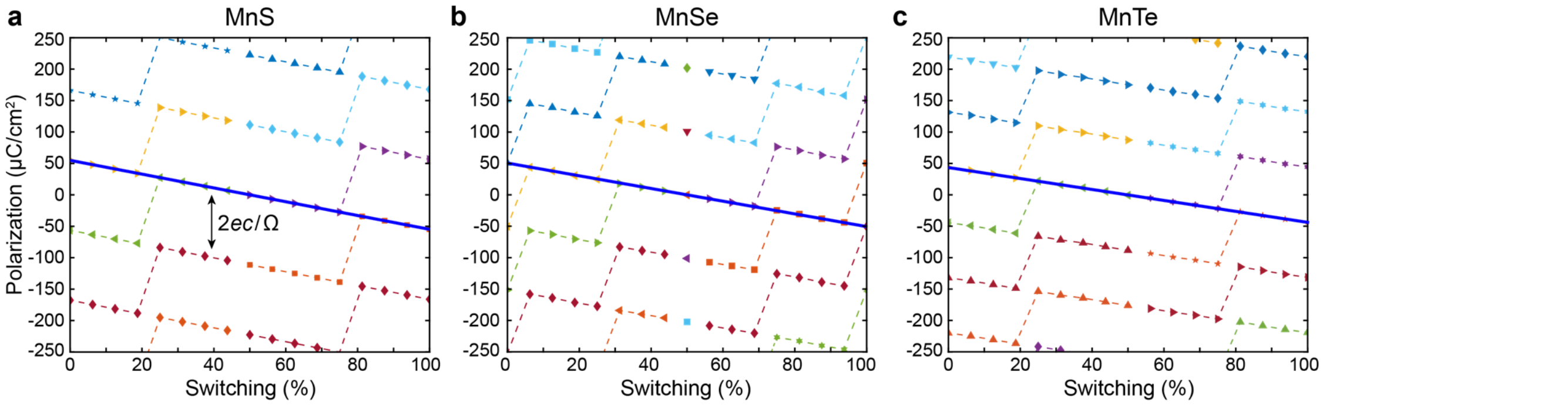

**FIG. D1**: The polarization evolution of (a) MnS, (b) MnSe, and (c) MnTe. The polarization difference between the end and start points of the polarization branches (guided by the solid blue curves) is the double spontaneous polarization. The fluctuations and color changes along the curves are due to numerical jumps arising from the multivalued nature of polarization. The magnitude between adjacent branches reflects the polarization quantum $2ec/\Omega$, indicated by the black arrow in panel (a).

The Born effective charges are calculated using density functional perturbation theory implemented in VASP [75]. The components $Z^*_{zz}$ for cations and anions are shown in Table D1. For MnS to MnTe, the Born effective charges are close to the nominal valence of 2, with a slight increase observed for lighter chalcogen elements, consistent with their higher electronegativity.

To validate our results of NEB calculations, we calculated the polarization switching energy barriers of benchmark wurtzite compounds. For comparison, we obtain switching barriers of 511 meV/f.u. for AlN and 249 meV/f.u. for ZnO. These values are consistent with previously reported results, which fall in the range of 500–700 meV/f.u. for AlN [ 76 , 77 , 78 ] and approximately 260 meV/f.u. for ZnO [79].

**Table D1.** Calculated electric polarization and Born effective charges for different wurtzite compounds.

| Compound | Polarization ($\mu C/cm^2$) | Born effective charge $Z^*_{zz}$ (e) |
|---|---|---|
| MnS | 54.9 | 2.045 |
| MnSe | 50.6 | 2.031 |
| MnTe | 43.4 | 2.023 |
| ZnO | 81.2 | 2.208 |
| AlN | 133.7 | 2.680 |

## APPENDIX E: SYMMETRY ALLOWED LINEAR AND NONLINEAR ELECTRONIC TRANSPORT TENSORS

The electric response current induced by external electric field up to second order is given by $j_a = \sigma_{b;a} E_b + \sigma_{bc;a} E_b E_c$ , where $a$, $b$, and $c$ denotes Cartesian components. The first order conductivity tensor $\sigma_{b;a}$ contains anomalous Hall effect (AHE) and conventional Drude contributions. The second order tensor $\sigma_{bc;a}$ includes contribution from the Berry curvature dipole (BCD) [70], quantum metric dipole (QMD) [71], and Drude terms. Table E1 summarizes the magnetic symmetry allowed conductivity tensors for each magnetic phase. Table E2 provides the relative signs of the allowed tensor components between the two opposite polarization states for both out-of-plane and in-plane magnetic moment orientations.

**Table E1.** Magnetic symmetry–allowed components of the linear and second-order conductivity tensors for the AIAO, $AFMs_a$, $AFMs_b$, and AM phases. The corresponding magnetic space groups (MSG) and magnetic point groups (MPG) are listed, together with the allowed linear anomalous Hall effect (AHE) components and second-order nonlinear responses associated with the Berry curvature dipole (BCD) and the quantum metric dipole (QMD). The underlying magnetic structure for each phase is depicted.

| Magnetic phase | MSG (MPG) | Linear AHE | BCD | QMD | Magnetic structure |
|---|---|---|---|---|---|
| AIAO | $P3m'1$ $(m'm'2)$ | $\sigma_{x;y} = -\sigma_{y;x}$ | $\sigma_{xx;z}, \sigma_{yy;z}, \sigma_{yz;y}, \sigma_{xz;x}$ | $\sigma_{xx;x}, \sigma_{yy;x}, \sigma_{xy;y}, \sigma_{yz;x}, \sigma_{xz;y}$ | 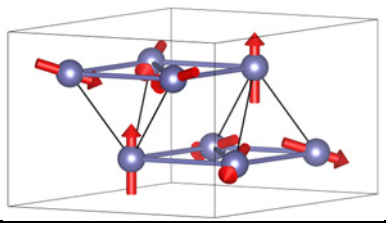 |
| $AFMs_a$ $\hat{m}\|\|[001]$ | $P_Cna2_1$ $(mm2.1')$ | – | $\sigma_{xx;z}, \sigma_{yy;z}, \sigma_{xz;x}, \sigma_{xz;y}, \sigma_{yz;x}, \sigma_{yz;y}, \sigma_{xy;z}$ | – | 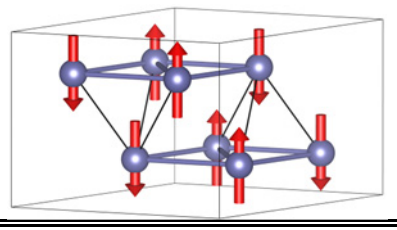 |
| AFMs$_a$ $\hat{m}\|\|[100]$ | $P_a2_1$ (2.1′) | – | $\sigma_{xx;z}$, $\sigma_{yy;z}$, $\sigma_{xz;x}$, $\sigma_{xz;y}$, $\sigma_{yz;x}$, $\sigma_{yz;y}$, $\sigma_{xy;z}$ | – | 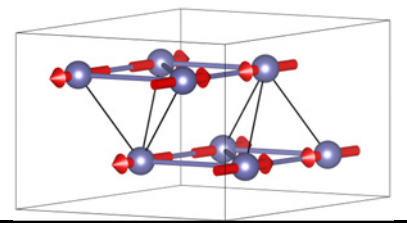 |
| AFMs$_a$ $\hat{m}\|\|[1\bar{1}0]$ | $P_Cca2_1$ $(mm2.1')$ | – | $\sigma_{xx;z}$, $\sigma_{yy;z}$, $\sigma_{xz;x}$, $\sigma_{xz;y}$, $\sigma_{yz;x}$, $\sigma_{yz;y}$, $\sigma_{xy;z}$ | – | 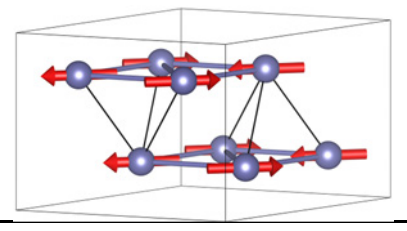 |
| AFMs$_b$ $\hat{m}\|\|[001]$ | $P_Cca2_1$ $(mm2.1')$ | – | $\sigma_{xx;z}$, $\sigma_{yy;z}$, $\sigma_{xz;x}$, $\sigma_{xz;y}$, $\sigma_{yz;x}$, $\sigma_{yz;y}$, $\sigma_{xy;z}$ | – | 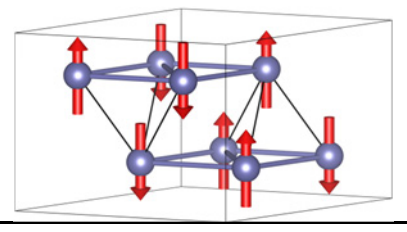 |
| AFMs$_b$ $\hat{m}\|\|[100]$ | $P_a2_1$ (2.1′) | – | $\sigma_{xx;z}$, $\sigma_{yy;z}$, $\sigma_{xz;x}$, $\sigma_{xz;y}$, $\sigma_{yz;x}$, $\sigma_{yz;y}$, $\sigma_{xy;z}$ | – | 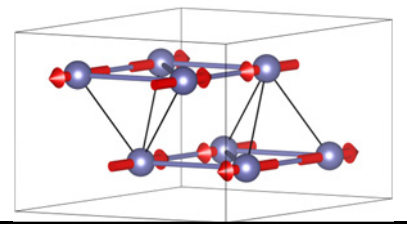 |
| AFMs$_b$ $\hat{m}\|\|[1\bar{1}0]$ | $P_Cna2_1$ $(mm2.1')$ | – | $\sigma_{xx;z}$, $\sigma_{yy;z}$, $\sigma_{xz;x}$, $\sigma_{xz;y}$, $\sigma_{yz;x}$, $\sigma_{yz;y}$, $\sigma_{xy;z}$ | – | 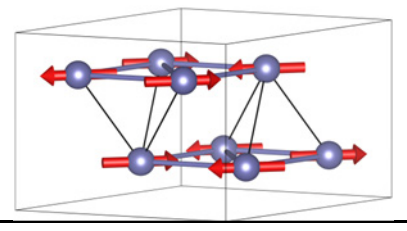 |
| AM $\hat{m}\|\|[001]$ | $P6'_3m'c$ $(6'mm')$ | – | $\sigma_{xx;z}$, $\sigma_{yy;z}$, $\sigma_{xz;x}$, $\sigma_{yz;y}$ | $\sigma_{xx;x}$, $\sigma_{yy;x}$, $\sigma_{xy;y}$ | 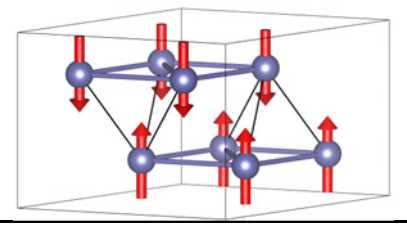 |
| AM $\hat{m}\|\|[100]$ | $Cmc2_1$ $(mm2.1)$ | – | $\sigma_{xx;z}$, $\sigma_{yy;z}$, $\sigma_{xz;x}$, $\sigma_{xz;y}$, $\sigma_{yz;x}$, $\sigma_{yz;y}$, $\sigma_{xy;z}$ | $\sigma_{yy;z}$, $\sigma_{xz;x}$, $\sigma_{xz;y}$, $\sigma_{yz;x}$, $\sigma_{yz;y}$ | 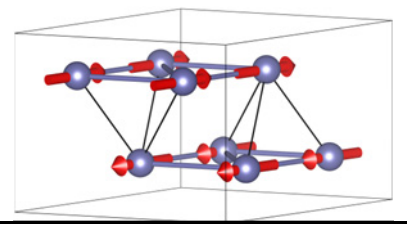 |
| AM $\hat{m}\|\|[1\bar{1}0]$ | $Cm'c'2_1$ $(m'm'2)$ | $\sigma_{x;y} = -\sigma_{y;x}$ | $\sigma_{xx;z}$, $\sigma_{yy;z}$, $\sigma_{xz;x}$, $\sigma_{xz;y}$, $\sigma_{yz;x}$, $\sigma_{yz;y}$, $\sigma_{xy;z}$ | $\sigma_{xx;z}$, $\sigma_{yy;z}$, $\sigma_{xz;x}$, $\sigma_{xz;y}$, $\sigma_{yz;x}$, $\sigma_{yz;y}$ | 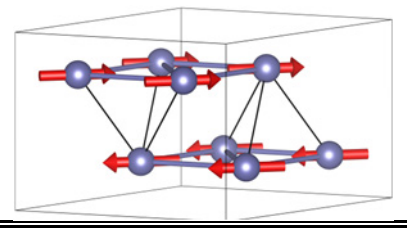 |

**Table E2.** Relative sign changes of the conductivity tensor components between opposite polarization states $(N, P)$ and $(N, -P)$ for out-of-plane and in-plane magnetic moment orientations, where $N$ denotes the Néel vector and $P$ the ferroelectric polarization. Different response types are listed: linear anomalous Hall effect (AHE), Berry curvature dipole (BCD), and quantum metric dipole (QMD). A preceding minus sign indicates a sign change of the corresponding tensor component upon polarization reversal.

| Symmetry relation between (*N*,*P*) and (*N*,–*P*) states | Response type | Effect of polarization reversal on response tensor components |
|---|---|---|
| 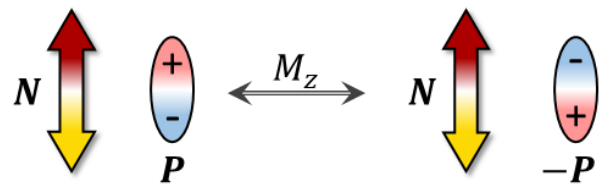 | Linear AHE | – |
| | BCD | $\sigma_{xx;z}$, $\sigma_{yy;z}$, $\sigma_{xz;x}$, $\sigma_{xz;y}$, $\sigma_{yz;x}$, $\sigma_{yz;y}$, $\sigma_{xy;z}$ |
| | QMD | $\sigma_{xx;x}$, $\sigma_{yy;x}$, $\sigma_{xy;y}$, $-\sigma_{xx;z}$, $-\sigma_{yy;z}$, $-\sigma_{xz;x}$, $-\sigma_{xz;y}$, $-\sigma_{yz;x}$, $-\sigma_{yz;y}$ |
| 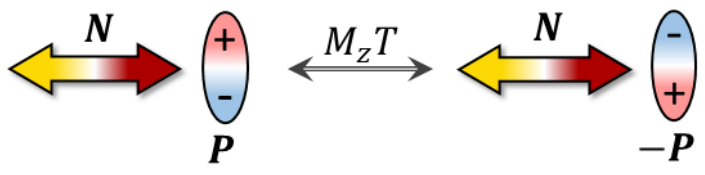 | Linear AHE | $-\sigma_{x;y}$ |
| | BCD | $\sigma_{xx;z}$, $\sigma_{yy;z}$, $\sigma_{xz;x}$, $\sigma_{xz;y}$, $\sigma_{yz;x}$, $\sigma_{yz;y}$, $\sigma_{xy;z}$ |
| | QMD | $-\sigma_{xx;x}$, $-\sigma_{yy;x}$, $-\sigma_{xy;y}$, $\sigma_{xx;z}$, $\sigma_{yy;z}$, $\sigma_{xz;x}$, $\sigma_{xz;y}$, $\sigma_{yz;x}$, $\sigma_{yz;y}$ |